# On the Constituent Attributes of Software and Organisational Resilience

Vincenzo De Florio[i]
Universiteit Antwerpen and iMinds research institute
Middelheimlaan 1, 2020 Antwerpen-Berchem, Belgium
e-mail: vincenzo.deflorio@ua.ac.be

# Abstract

Our societies are increasingly dependent on the services supplied by our computers and their software. Forthcoming new technology is only exacerbating this dependence by increasing the number, the performance, and the degree of autonomy and inter-connectivity of software-empowered computers and cyber-physical "things", which translates into unprecedented scenarios of interdependence. As a consequence, guaranteeing the persistence-of-identity of individual and collective software systems and software-backed organisations becomes an increasingly important prerequisite towards sustaining the safety, security, and quality of the computer services supporting human societies. Resilience is the term used to refer to the ability of a system to retain its functional and non-functional identity. In the present article we conjecture that a better understanding of resilience may be reached by decomposing it into a number of ancillary constituent properties, the same way as a better insight in system dependability was obtained by breaking it down into safety, availability, reliability, and other sub-properties. Three of the main sub-properties of resilience proposed here refer respectively to the ability to perceive environmental changes; to understand the implications introduced by those changes; and to plan and enact adjustments intended to improve the system-environment fit. A fourth property characterises the way the above abilities manifest themselves in computer systems. The four properties are then analyzed in three families of case studies, each consisting of three software systems that embed different resilience methods. Our major conclusion is that reasoning in terms of our resilience sub-properties may help revealing the characteristics—and in particular the limitations—of classic methods and tools meant to achieve system and organisational resilience. We conclude by suggesting that our method may prelude to meta-resilient systems—systems, that is, able to adjust optimally their own resilience with respect to changing environmental conditions.

# 1 Introduction

Computer systems are not dissimilar from critical infrastructures in which different mutually dependent components—in fact, infrastructures themselves—contribute to the emergence of an intended service. Thus, in computer systems the software infrastructure relies on the quality of the hardware infrastructure, and vice-versa a hypothetically perfect hardware would not result in the intended service without a corresponding healthy software infrastructure. Software resilience refers to the robustness of the software infrastructure and may be defined as the trustworthiness of a software system to adapt itself so as to absorb and tolerate the consequences of failures, attacks, and changes within and without the system boundaries. As a resilient body is one that "subjected to an external force is able to recover its size and shape, following deformation" (Harris, 2005), likewise software is said to be resilient when it is able to recover its functional and non-functional characteristics—its "identity"—following failures, attacks, and environmental changes. As critical infrastructures call for organisational resilience, likewise mission- and business-critical computer systems call for software resilience. Understanding and mastering software resilience is a key prerequisite towards being able to design effective services for complex and ever changing deployment environments such as those characterizing, e.g., ubiquitous and pervasive environments.

In what follows we consider resilience as a collective property, i.e., a property better captured by considering a number of sub-properties, each of which focuses on a particular aspect of the whole. This method was successfully used by Laprie and others to better characterize dependability as a collective property described by constituent attributes—availability, reliability, safety, maintainability, and others (Laprie, 1985; Laprie, 1995). A major result of applying this decomposition method is that it provides us with a "base" of attributes with which one may describe—and limit—the characteristics of existing methods, systems, and algorithms. Thanks to Laprie's efforts we now can more precisely describe the behaviours of a computer entity as being, e.g., reliable, but not available; or safe, but not reliable; and so forth. In turn, this makes it more apparent whether a certain solution matches an intended mission or a given hypothesised environment. This also helps reasoning about the consequences of erroneous deployments or those related to the drifting between the system assumptions and the actual environmental conditions (De Florio, 2010).

Aim of the current paper is to apply to resilience the same method that Laprie and others applied to dependability. Our conjecture and hope is that similar insight may ensue from this, and that our sub-properties may prove to constitute an effective resilience *base*, namely a set of independent and complementary variables useful to reason about the qualities and the shortcomings of a given resilient entity with respect to its deployment conditions.

The rest of the current paper is structured as follows: In Section 2 we define the terms and introduce the main method of our discussion and our base of orthogonal sub-properties of resilience.
In Section 3 we use our base of sub-properties to describe and characterise three families of systems. Each family consists of a non-resilient member and two adaptive or resilient reformulations from our current and past research activity. In this section we also show how discussing the resilience of a (software) system in terms of its constituent properties helps exposing several important aspects of those systems, including their complexity, predictability, and dependence on design assumptions. We conjecture that this may provide designers with a convenient tool to reason about the cost-effectiveness of resilient methods and approaches. In turn, this feature may pave the way towards future autonomic meta-resilient systems—namely systems able to self-optimise their own resilience with respect to variable environmental conditions.
Our conclusions are finally drawn in Section 4.

# 2 Method

In this section we introduce the main terms and the method of our discussion. Our starting point is behavioural and extends the classic works of Rosenblueth (Rosenblueth *et al.*, 1943) and Boulding (Boulding, 1956). The focus here is not on general systems behaviour but rather on resilient behaviours—those behaviours that are meant to guarantee the functional and non-functional identity of the system at hand.
In what follows we first define resilience and then characterise four classes of behaviours that are typical of resilient systems. Finally, we present four independent properties that—we conjecture—collectively describe the major aspects of resilience.

## 2.1 Resilience

Resilience is defined in (Meyer, 2009) as the ability to tolerate (or even profit from) the onset of unanticipated changes and environmental conditions that might otherwise cause a loss of acceptable service. Jen (Jen, 2004) characterises in more detail resilience as a measure of a system's persistence of both functional and non-functional features under specified and unforeseen perturbations. By analyzing the above definitions one may further detail resilience as the ability to enact and trade-off between two main behaviours:

1. Continuously re-adjusting oneself so as to improve the system-environment fit and compensate for foreseen and/or unpredicted changes in the environment.
2. Making sure that said changes and adjustments do not affect the identity of the system — in other words, its peculiar and distinctive functional and non-functional features.

We believe that the above two behaviours closely correspond to the Aristotelian concept of *entelechy* (Aristotle, 1986), a complex property capturing an entity's ability to pursue "completion" (that is, one's optimal state) without compromising the specific nature and aspects that define itself. The relationship between resilience and entelechy is more evident when considering the ingenious translation of entelechy proposed by Sachs (1995) as the property of "being-at-work-staying-the-same": in fact "being at work" may be interpreted as behavior (1.) while "staying the same" corresponds to behavior (2.)

Obviously resilience is an important prerequisite to a system's correct and effective operation, which explains the ever increasing attention devoted to such property in recent times. We now describe four main methods that are used to achieve resilience.

## 2.2 Resilient behaviours

Rosenblueth, Wiener, and Bigelow (*Rosenblueth et al.*, 1943) proposed a classification of systems according to their behaviour and purpose, which they defined respectively as "any change of an entity with respect to its surroundings" and "awareness of voluntary activity". In particular in their cited work the author developed a hierarchy consisting of the following behavioural classes:

1. Systems characterised by passive behaviour: no source of "output energy" (Rosenblueth *et al.*, 1943) may be identified in any activity of the system.

2. Systems with active, but non-purposeful behavior—systems, that is, that do not have a "specific final condition toward which they strive."
3. Systems with purposeful, but non-teleological (i.e., feedback-free) behaviour: systems, that is, in which "there are no signals from the goal which modify the activity of the object" (*viz.*, the system) "in the course of the behaviour."
4. System with teleological, but non-extrapolative behaviour: in other words, systems that are purposeful but are unable to construct models and predictions of a future state to base their reactions upon.
5. First-order predictive systems, namely systems able to extrapolate along a single perception dimension.
6. Higher-order predictive systems, or in other words systems that are able to base their reactions on the correlation of two or more perception dimensions, possibly of different nature—temporal and spatial coordinates, for instance.

A seventh class was introduced later by Kenneth Boulding in his classic article on General Systems Theory (Boulding, 1956):

7. Collective adaptive systems, e.g. (digital or natural) ecosystems, socio-technical systems, multi-agent systems, social organisations (De Florio & Blondia, 2010), and cyber-physical societies (Zhuge, 2010). Such systems are characterised by complex *collective* behaviours resulting from the sum of possibly many and variegated individual behaviours such as those in (1.)—(6.)

From the point of view of resilient behaviours, we observe what follows:

- Passive-behaviour systems (1.) and active, non-purposeful behavior systems (2.) may only achieve simple forms of resilience through prior arranged **redundancy**. Such systems are in fact **close-world**, in that they do not require any form of **perception** of physical world properties. Due to their peculiar design assumptions, systems in this category do not require any form of **awareness** nor they do need any ability to **plan** reactions or countermeasures to external events. The term we use to refer to systems in this category is Ataraxies[ii]. Ataraxies reach resilience through **elasticity**, namely "redundancy of their physical resources such as plant and machinery, locations or buildings, and the lifelines infrastructure on which they rely" (Stephenson et al., 2010). A strong dependency on system and environmental assumptions characterises Ataraxies, which obviously translates in inherent fragility with respect to drifting or failures of those assumptions (De Florio, 2011c).
- Purposeful-only (3.) and teleological behaviour systems (4.) usually attain resilience through **event masking**, obtained in turn through redundancy and voting mechanisms (De Florio et al., 1998). Special cases of this strategy are, e.g., fault masking, whose classical application in software systems is known as *N*-version programming (Avižienis, 1985), and data integrity assurance methods—used, for example, in redundant data structures and in other methods to mask disturbances due to interference or attacks. A noteworthy case of event masking is given by systems that exhibit a spontaneous emergence of modularity—a phenomenon that is known to enhance the resilience of network and systems (Clune et al., 2013). Systems in this category require limited perception and awareness and are usually characterised by predefined reactions. "Thermostats" and "Cells" are the terms used by Boulding (1956) to refer respectively to class (3.) and class (4.) entities. "Servo-mechanisms" is the term used in (Rosenblueth et al., 1943) to refer to systems in (3.)
- Predictive systems—both simple (5.) and complex (6.)—attain resilience through **parametric and structural adaptation**. In the first case the systems do not modify their structure and only tune their workflows; structural adaptation on the contrary involves a topological revision of the system and its workflow. In both cases resilience calls for several complex features: (a) some form of perception; (b) complex diagnosis of the situations at hand, including (b.1) introspection capabilities able to reveal conditions and situations originating within and without the system boundaries and threatening the correct execution; (b.2) the ability to compare current and past situations, to unravel similarities and trends, and to identify causes of faults, errors, and losses of identity (De Florio, 2012a); (c) the ability to plan reactive or proactive strategies to compensate for current (or, respectively, future) threatening conditions; and (d) the ability to enact those strategies by performing the actual parametric and structural adaptations. The above features are often referred to as the phases (or steps) of a MAPE ("Monitor, Analysis, Planning, and Execution") control loop (Kephart & Chess, 2003). We shall refer to this class of methods as to **predictive individual behaviours**. "Plants", "Animals", and "Human Beings" are corresponding classes in the classification of (Boulding, 1956), the main difference among them being the degree of complexity of the above features and in particular of (c).
- Collective adaptive systems may range from groups of cells and colonies of bacteria to complex social organisations such as digital, knowledge, and business ecosystems. Due to such a profound diversity, *individual* behaviours may range from the simplest to the most complex and organized (Guseva, 2012) (Schultz et al., 2009). The emerging **resilient social behaviours**, though, are often characterised by high complexity and sophisticated features (Astley & Fombrun, 1983) (Adner & Kapoor, 2010) (Brandenburger & Nalebuff, 1998). Resilience of the species; emergence of self-similarity, modularity, and fractal dimensions; and the ability to reach a harmony or dynamic equilibrium with the deployment environment, are typical traits of said social behaviours (Webb & Levin, 2004), (De Florio et al., 2013). Though complex perception capabilities are often a strong prerequisite to resilient social behaviours, awareness and planning may range from the simple forms of cellular and bacterial life forms to highly sophisticated forms such as those characterizing business ecosystems, e.g., co-evolution (Adner & Kapoor, 2010) and co-opetition (Brandenburger & Nalebuff, 1998).

Elasticity, event masking, resilient individual and social behaviours represent four main methods that may be applied to achieve resilience in systems and organisations. As mentioned above, each of these methods call for a different "blend" of abilities and features, which are summarized in Table 1.

Table 1. Resilient methods vs. resilient requirements and properties

|  | Redundancy | **Perception** | **Awareness** | **Planning** |
|---|---|---|---|---|
| **Elasticity** | Yes | None | None | None |
| **Event masking** | Yes | None or Limited | None or limited | None |
| **Predictive (individual) behaviours** | Yes | Limited-to-abundant | Moderate-to-abundant | Limited-to-abundant |
| **Resilient (social) behaviours** | Yes | Moderate-to-abundant | Limited-to-abundant | Limited-to-abundant |

It is important to remark that the above classifications of resilient behaviour and resilient methods do not represent an assessment of appropriateness, let alone of the "quality", of the corresponding entities. Elasticity, for instance, may be the best choice when resources are abundant but the system at hand is a close-world entity with little or no perception, awareness, and planning capability. In other words, an appropriate choice requires that both the characteristics of the environment and the mission requirements match at the same time the available resources and the characteristics of the players. Furthermore, complex characteristics are often associated to higher costs and, what is even more important, to a larger amount of complexity. In turn, the higher a system's complexity, the more difficult it is to prove that the system still complies with its functional and non-functional specifications.

Perception, Awareness, and Planning, are three behavioural attributes that, we conjecture, can be effectively used to describe and characterise the above mentioned generic resilient *methods*. On the other hand, when applied to describe specific real-life *systems* together with their concrete implementation, we found it convenient to complement those properties with a fourth one. This fourth ancillary constituent, which we shall call Dynamicity, represents another independent "dimension" for the characterisation of a system's or an organisation's resilience. Given a Perception, Awareness, or Planning sub-system, Dynamicity will be used in what follows to express one of the following three attributes:

1. (Low Dynamicity) The sub-system is statically selected at design-time and impossible to change while operating.
2. (Medium Dynamicity) The sub-system may be dynamically selected or unselected at run-time.
3. (High Dynamicity) The sub-system is the dynamic product of a collective strategy run by a plurality of system components.

The following notation will be used to express that resilience sub-system $p$ of system $s$ exhibits class-$b$ behaviours and class-$d$ Dynamicity: $s.p \rightarrow (b, d)$. As an example, if the Awareness sub-system of $s$ is characterised by predictive behaviour that cannot be revised at run-time we shall say

$$s.\text{Awareness} \rightarrow (\text{predictive}, \text{low}).$$

In order to discuss and qualitatively evaluate our proposition we now present three families of systems and make use of Perception, Awareness, Planning, and Dynamicity to characterise their resilience.

# 3 Perception, Awareness, Planning, and Dynamicity in Systems and Organisations

Perception, Awareness, and Planning, together with their Dynamicity, will be used in this section as a way to characterize resilience in three families of system. The first such family is given by three Modular Redundancy systems and is discussed in Section 3.1. The major diversifying factor will be in this case Awareness. A second family of adaptive and meta-adaptive systems is then presented in Section 3.2. In this second case Planning and Dynamicity play a key role. Finally, a family of social organisations is discussed in Section 3.3. The accent here is on collective forms of Perception and Awareness and their resulting Dynamicity. In each section we also highlight other useful aspects and specific characteristics of the employed Perception, Awareness, and Planning components, namely their algorithmic complexity (how computationally intensive a given component is); predictability (how difficult it is to anticipate the action of the component); and dependence on assumptions (how sensitive the component is to the drifting or the invalidation of the system's design assumptions.)

## 3.1 Modular Redundancy Systems

Modular redundancy is a well-known strategy to mask defects and disturbances ranging from physical to design faults as well as environmental interference both in hardware and software (De Florio & Blondia, 2008b). In software modular redundancy calls for specific requirements, in that the occurrence of design faults is likely to induce correlated failures. Design diversity is a methodology to deal with this problem and N-version programming (NVP) is a practical approach to enact it (Avizienis, 1985). Both in hardware and in software, modular redundancy is based on a simple mechanism:

1. An input value is multiplexed onto a set of replicas (called "versions" in NVP).
2. All replicas process independently the input value and forward their output to an adjudicator module.
3. The adjudicator received the replicas' output values and chooses the system output.

Often the adjudicator is implemented as a voter that partitions the set of output values into equivalence classes and determines the system output according to simple rules, for instance the presence of an equivalence class consisting of a majority of output values. More complex strategies exist but, for the sake of simplicity, in what follows we shall assume the just mentioned voting strategy to hold. Under the above assumptions, modular redundancy based on $2N+1$ replicas allows for up to $N$ defects or disturbances to be masked.

A key aspect in the design of a modular redundancy system is the choice of $N$. Such value represents the worst case scenario that the system is supposed to experience. Typically this is the result of worst-case analyses and observations of the system and its target deployment environment.

A model of the cumulative effect of the system defects and the deployment environment is given by dynamic system $cr_{(S,E)}(t)$ defined as follows:

> $cr_{(S,E)}(t)$ is the minimum number of replicas necessary to withstand the faults or disturbances experienced at time $t$ by system $S$ when deployed in environment $E$.

In the absence of faults or disturbances, at time $t$, $cr_{(S,E)}(t)$ is equal to 1: a single replica suffices to fulfill the mission and guarantee the identity of the system. In the face of faults or disturbances affecting $N$ replicas at time $t$, $cr_{(S,E)}(t)$ is equal to $2N+1$, as this implies that there exists a minimum majority of $N+1$ non-affected replicas in $S$. We shall refer to function $cr_{(S,E)}$ as to the "contextual redundancy function" and denote it as $cr$ provided that this can be done without introducing ambiguity.

$N$ and $cr$ statically define a "Resilience Region" (RR) as well as an "Unresilience Region" ($\neg$RR): if for some $t$ contextual redundancy is greater than $N$, then the system shall experience in $t$ a loss of identity (in this case, a failure) despite its fault-tolerant design.

It is important to highlight here another important aspect related to the operational costs of the system. In fact, from the cost perspective, $N$ and the contextual redundancy function define two other regions of interest:

- An Undershooting Region ($\vee$R), in which the system's foreseen redundancy, interpreted here as a measure of the cost of the system, proves to be insufficient to counteract the current faults and disturbances (i.e., $N <$

$cr(t)$). Quantity $\vee(t) = cr(t) - N$ represents a measure of the discrepancy between the worst-case analysis of the designer and what is actually experienced by the system at time $t$.
- An Overshooting Region ($\wedge R$), safely part of the Resilience Region though overabundant with respect to the currently experienced threat. Metric $\wedge(t) = N - cr(t)$ quantifies the cost in excess (in terms of redundancy) paid by the system at time $t$ to preserve its identity.

Due to the dynamic nature of $cr(t)$, both $\vee R$ and $\wedge R$ drift continuously in time.

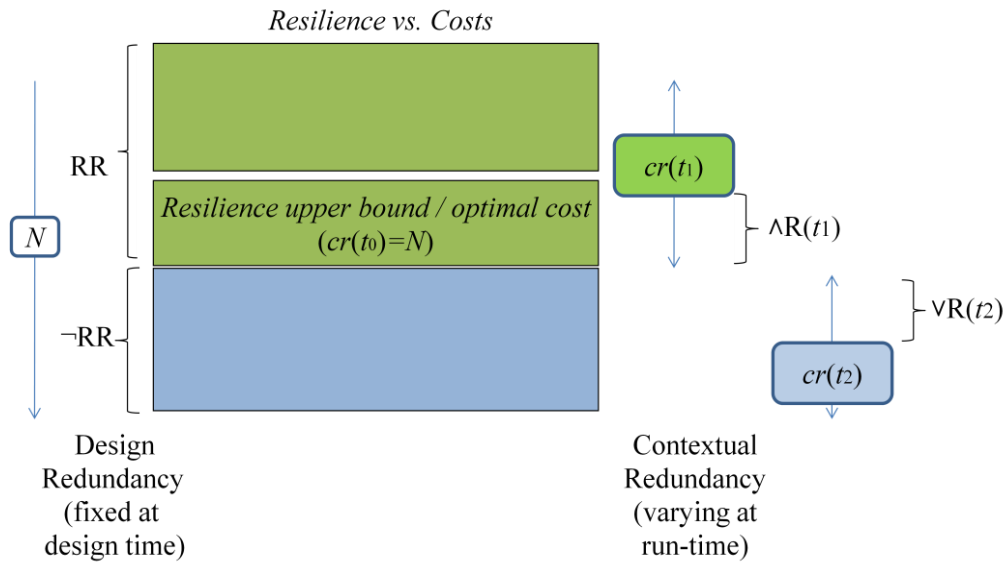

Figure 1: Pictorial representation of $N$, contextual redundancy, and the four regions RR, ¬RR, $\wedge R$ (at time $t_1$, in which a majority is found), and $\vee R$ (at time $t_2$, in which no majority can be found). Light green means resilient, blue is used for unresilience. Note how, at time $t_0$, the system reaches its maximum efficiency (and $\wedge R(t_0) = \vee R(t_0) = \emptyset$).

Figure 1 exemplifies the just defined four regions.

In the rest of the current section we introduce a resilient system and two reformulations. By means of the resilient properties and methods introduced earlier we discuss the positive and negative aspects of said systems with respect to resilience and other metrics.

### 3.1.1 Redundant Data Structures

A system based on modular redundancy is given by so-called Redundant Data Structures (RDS) (De Florio & Blondia, 2008a). The design goal of RDS is to guarantee data integrity against environmental disturbances affecting a computer system's memory modules. When such disturbances only affect single memory cells and do not induce more catastrophic failures (as described, e.g., in (De Florio, 2010)), then RDS may effectively protect against, e.g., electro-magnetic interference. As already mentioned, RDS specialises modular redundancy for data integrity and is based on the following simple strategy:

1. Memory is organized as an array of "replicated cells". Each replicated cell **c** consists of $2N+1$ "conventional" memory cells, $\{c_0, c_1, \ldots, c_{2N}\}$, organized according to some strategy[iii].
2. Writing value $v$ into replicated cell **c** means writing $v$ on each of the $c_i$, $i = 0,\ldots,2N$.
3. Reading a value from replicated cell **c** means performing majority voting on the current value of the $c_i$, $i = 0,\ldots,2N$.
4. Optionally, values read from replicated cell **c** may be used to "refresh" the $c_i$, thus rejuvenating all conventional memory cells associated to **c**.

By considering the metrics defined in previous section, we observe what follows:

- RDS is a simple **event masking** mechanism.
- RDS is a very simple **purposeful-behaviour** system—with the terminology introduced in previous section, RDS is a servo-mechanism (Rosenblueth *et al.*, 1943) or a "Cell" (Boulding, 1956).
- **Perception** is absent: RDS is an Ataraxy in that it only perceives physical events through the effects those events have on the memory cells.

- **Awareness** is purely **purposeful**. The only Awareness reached by RDS consists of assessing whether a majority may be adjudicated or not.
- **Planning** is also simple and **purposeful**.
- **Dynamicity** is absent, in that all mechanisms are simple and statically predefined at design time.

Other considerations are as follows:

- Resilience depends on the truth value of design assumptions and in particular on the choice of *N*.
- The choice of *N* represents the result of some worst-case analysis that refers to the conditions of a reference deployment environment as manifested during a certain interval of time. A "hidden design assumption" here (in the sense clarified in (De Florio, 2010)) is that the drifting and the evolution of such environment will not lead to undershootings. No check or self-check is usually foreseen to test the validity of this assumption, which led in the past to well-known and appalling accidents as, e.g., the Therac-25 malfunctions.
- The algorithmic complexity introduced by RDS is very limited and statically predefined and known at design time.
- The operational cost of the system (in terms of redundancy and energy requirements) is predefined and known at design time.

We remark how, on the one hand, in the absence of assumption failures the above assumptions imply that RDS is a predictably stable and resilient system. On the other hand, the limited Perception and Awareness, as well as the hidden dependence on the deployment environment, make of RDS an "easy victim" for any change affecting the system and the environmental assumptions. By means of a well-known American vernacular we could say that RDS systems are "sitting ducks" with respect to those changes[iv].

## 3.1.2 Adaptively-redundant Data Structures

In (De Florio and Blondia, 2008) we introduced a reformulation of RDS that we call Adaptively-redundant Data Structures (A-RDS). The idea behind A-RDS is quite simple: instead of a fixed value for *N*, A-RDS uses $N(t)$ replicas, where $N(t)$ is a dynamic system that tries to minimize overshootings while guaranteeing that the system stays in the Resilience Region at all times. Such teleological, predictive behavior is reached through the following procedure:

1. Memory is organized as an array of "replicated cells". Each replicated cell **c** consists of $2N(t)+1$ "conventional" memory cells, $\{c_0, c_1, …, c_{2N(t)}\}$, organized as with RDS. $N(0)$ is set to some *N* estimated as it was done in the case of RDS.
2. Writing value *v* into replicated cell **c** means writing *v* on each of the $c_i$, $i = 0,…,2N(t)$.
3. Reading a value from replicated cell c means performing majority voting on the current value of the $c_i$, $i = 0,…,2N(t)$.
4. After each successful majority voting, which we assume took place at time $t_{prev}$, a so-called "distance-to-failure" function (DTOF) is invoked, defined as follows:

$$\text{DTOF}(t_{now}) = \wedge(t_{prev}) / N(t_{prev}).$$

   DTOF provides an estimation of the risk of undershooting at time $t_{now}$ as the ratio of the overshooting experienced at time $t_{prev}$ with the redundancy of the system at time $t_{prev}$. DTOF ranges from 0 (minimal distance to failure, thus maximal risk of a loss of identity) to 1 (maximal distance to failure, viz. minimal risk). When $\text{DTOF}(t_{now})$ is less than a given threshold, $N(t_{now})$ is increased by 2 units, namely two additional replicas are added to the system.
5. If DTOF stays beyond the threshold for a given amount of time, then $N(t_{now})$ is decreased by 2.
6. If a majority voting fails, $N(t_{now})$ is increased by 2 units.

As we have done in last section for RDS, here we state a number of remarks based on the metrics defined in Section 2:

- A-RDS is a **first-order predictive mechanism** that attains resilience through **parametric adaptation**. With the terminology introduced in Section 2, A-RDS is a "Plant."
- **Perception** is absent. A-RDS is another example of Ataraxy: as it was the case for RDS, also A-RDS only perceives physical events through the effect those events have on the memory cells.
- **Awareness** is **first-order predictive**. It is reached through introspection by assessing a risk of identity loss by considering a single and simple metric.
- **Planning** is simple and **non-teleological purposeful**.

- **Dynamicity** is low, as all mechanisms are statically predefined at design time.

Again we state a few additional remarks.
- Resilience depends on the ability of the awareness method to respond to the actually experienced threats. If the estimation of the risk of undershooting at time $t_{now}$ is correct, A-RDS is able to adjust its key parameter, $N(t)$, in such a way as to guarantee resilience while considerably reducing the operational cost of the system. This corresponds to environmental scenarios that change smoothly and continuously. In the face of more turbulent environment, A-RDS experiences considerable resilience failures (De Florio, 2011b).
- The algorithmic complexity introduced by A-RDS is greater than the one characterizing RDS. In particular the parametric adaptation of $N(t)$ is a costly step as it requires a reallocation of the whole set of redundant variables.
- The operational cost of the system (in terms of redundancy and energy requirements) varies with $N(t)$ and with the number of parametric adaptations enacted by A-RDS. Only a predefined upper bound for $N(t)$ is known at design time.

We remark how the above assumptions imply that A-RDS is considerably *less predictable* than RDS. On the other hand, the dependence on the deployment environment is explicit. In particular resilience failures are easily detected as undershootings, the frequency of which may be interpreted as a quantitative measure of the turbulence of the deployment environment: the more A-RDS is tricked into a misprediction, the more the environment is likely to be discontinuous and unpredictable.

### 3.1.3 Adaptive *N*-version Programming

A second reformulation of RDS is given through Adaptive *N*-version Programming (A-NVP), an approach and software architecture that further expands the features of A-RDS. A-NVP differs from A-RDS in several aspects: its primary design goal is not specifically data integrity and protection from faults and interference affecting computer memories—as it was the case for A-RDS—but rather providing an adaptive reformulation of *N*-version programming. A-NVP is also based on a voting mechanism, but the "votes" are not values read from main memory but the outputs produced by a number of software programs, called versions. The system employs $N(t)$ such versions and uses improved versions of DTOF to adapt such parameter. A-NVP also adopts structural adaptation as it keeps track of the performance of all active versions available. Through a mechanism of gradual penalisation and reward, all available versions are ranked by comparing their output with the one obtained through majority voting. This allows, at any time $t$, the top ranked $N(t)$ versions to be selected. The just sketched optimal selection is very important in that modular redundancy systems are known to fulfill their design goals only provided that their constituents exhibit at least a minimal amount of quality themselves[v] (Johnson, 1989). A-NVP is based on Web Services. A full discussion of A-NVP is out of the scope of this paper and may be acquired in (Buys et al., 2011) and (Buys et al., 2012).

Again a number of remarks derive by considering the metrics defined in Section 2:

- The above concise description already allows A-NVP to be identified as a **second-order predictive mechanism**, in that its behavior is a response computed by correlating two perception dimensions: the overall distance-to-failure (introduced in Sect. 3.1.2) as well as the versions' trustworthiness with respect to the majority of votes. With the terminology introduced in Section 2, A-NVP is an "Animal."
- A-NVP achieves resilience through **parametric and structural adaptation**, that is, by adjusting $N(t)$ and by replacing the versions so as to select the best $N(t)$ candidates available.
- **Perception** is absent: despite its more advanced characteristics with respect to RDS and A-RDS, still A-NVP is **quasi-close-world** as it also perceives physical events only through the effect those events have on the versions.
- **Awareness** is **second-order predictive** in that the risk of identity loss is assessed by considering two metrics at the same time.
- **Planning** is typically **non-teleological purposeful**.
- **Dynamicity** is low for Perception and Awareness as the corresponding requirements and algorithms do not allow dynamic revisions. Dynamicity is medium for Planning as it may be revised at run-time also because of the characteristics of the A-NVP software architecture, namely Web Services.

Other more general considerations follow:
- As in A-RDS, also in A-NVP resilience is mainly a result of the ability of the Awareness method to capture the nature of the threats manifested by the deployment environment. Preliminary experimentation of A-NVP indicates its ability to adjust and respond efficaciously to less stable environments than those matching the characteristics of A-RDS.
- The algorithmic complexity introduced by A-NVP is greater than the one characterizing RDS but considerably less than the one of A-RDS, mainly because the parametric adaptation of $N(t)$ is not as costly as in A-RDS in that it does not require heavy reorganisations of memory. The adoption of Web Services standards such as WS-ServiceGroup makes most of such complexity transparent to the programmer. On the other hand Web Services imply a best-effort approach that is detrimental to predictability.
- At the time of writing this paper we still lack a thorough analysis of the operational cost of the system, though it is sensible to expect a dependence of some nature on both $N(t)$ and the number of parametric and structural adaptations undergone. The scalability and the predictability of the system are also still under study.

### 3.1.4 Preliminary Conclusions

We have briefly introduced three modular redundancy systems. Here we briefly summarize their characteristics by means of the base of properties introduced in Sect. 2 and three ancillary and correlated properties.

Table 2. The three systems sketched in this section are listed in the first column, while corresponding (Behaviour, Dynamicity) couples are listed in column 2, 3, and 4. As an example, the element at row 3, column 4 states that A-RDS.Planning → (Purposeful, Low). Other ancillary properties are also summarised.

|       | Perception → | Awareness → | Planning → | Complexity | Predictability | Assumptions Dependence |
|-------|--------------|-------------|------------|------------|----------------|------------------------|
| RDS   | (Absent, Low) | (Purposeful, Low) | (Purposeful, Low) | Low | High | Very strong |
| A-RDS | (Absent, Low) | ($1^{st}$ order predictive, Low) | (Purposeful, Low) | Medium | Low | Strong |
| A-NVP | (Absent, Low) | ($2^{nd}$ order predictive, Low) | (Purposeful, Medium) | High | Medium | Medium? |

All systems are characterized by the absence of **Perception**. Minor differences concern **Planning**, while **Awareness** is where the major differences are. Such differences in fact classify the surveyed systems in three different behavioural and systemic categories. A direct consequence of this can be seen in the increasing Complexity of the three systems. Predictability is high in RDS due to its simple and stable algorithm (Jen, 2004), while it is low in A-RDS (mainly because of the high operational costs of memory reorganisation) and medium in A-NVP (due to the adopted best-effort approach). Assumptions dependence is where the introduced complexity pays off more, as the three systems appear to be characterised by decreasing dependence.

## 3.2 Adaptation and Meta-adaptation Systems

The systems described in Sect. 3.1 diversified themselves in their way to achieve resilience primarily through different Awareness capabilities. In this section instead we consider three other "design templates" for resilient systems with a special focus on the Planning step.

- The first case is given by the classic autonomic system realised through a so-called MAPE or MAPE-K control loop. In this case a single, predefined Planning module is considered.
- ACCADA: a system that allows the run-time context-aware and autonomic selection of a Planning module best matching the current systemic and environmental conditions.
- Transformer: an evolution of ACCADA that allows multiple plans issued by a set of contextually selected Planning modules to be merged together and executed during the run-time.

### 3.2.1 Simple MAPE loop systems

In their classical paper (Kephart & Chess, 2003) the authors introduce a general design template for autonomic systems, namely systems able to autonomously manage, organize, and in general, control, their action taking into account the current "context", namely the actual condition of the system, its environment, and its users. The proposed solution makes use of a so-called MAPE, where "M" stands for "Monitoring", "A" for "Analysing", "P" for "Planning", and "E" for "Executing". A MAPE control loop is a design template for extrapolative behaviour systems that: autonomously perceive change; analyse the impact of said change; plan a corresponding response; and execute said response. Most of adaptive and self-adaptive systems are realized through a MAPE loop or a specialisation thereof (Salehie & Tahvildari, 2009). MAPE loop may be concisely expressed in the application layer of conventional programming languages through mechanisms such as so-called reflective-and-refractive variables (De Florio & Blondia, 2007).

By considering the metrics defined in previous section, we observe what follows:

- MAPE loop is a general scheme to implement **parametric and structural adaptation systems**, namely systems based on **predictive individual behaviours**.
- MAPE loop systems are context-aware, hence Perception and Awareness play a key role. *A fortiori*, compliant systems are open-world.
- Specific and explicit support for a wide range of **Perception**, **Analysis**, and **Planning** behaviours is foreseen. No limitation is introduced by the scheme *per se* but, once the scheme is embedded in a computer system, resource limitations require a trade-off to be drawn by the system designer between specific M-, A-, P-, and E-subsystems (Nilsson, 2008; Nilsson, 2010).
- Usually pure MAPE loop systems do not foresee the possibility to adjust their own MAPE loop—in other words, they are not meta-adaptive (namely, able to adjust their own adaptation logics). This results in low **Dynamicity** and a strong dependency with the deployment environment. For the same reason this also results in a comparably simpler design and consequently in more stable systems (Jen, 2004) whose resilience may be more easily assessed and enforced.
- Resilience is to be managed by considering—reactively or proactively—the impact that both changes and adaptations have on the identity of the system.

We are now going to present two MAPE-loop systems that provide additional features and in particular meta-adaptive support to service-component applications.

### 3.2.2 ACCADA

ACCADA is a software framework for service-components MAPE-loop applications (Gui et al., 2011a; Gui et al., 2011b). ACCADA stands for "A Continuous Context-Aware Deployment and Adaptation framework", which concisely expresses the peculiar difference of such system with respect to pure MAPE-loop applications. The major benefit of ACCADA lies in fact in its support of run-time recomposition of all the application service-components, including those for managing adaptation, which are selected and de-selected autonomously. ACCADA assumes that a number of service-components are accessible through a registry and that said elements declare in meta-data their reference contextual conditions; their dependences; and their requirements. ACCADA is based on OSGi release 4 (Anonymous 2009). ACCADA responds to environmental and internal changes by performing a meta-adaptation procedure:

- First, a so-called Structural Modeller checks which of the available service-components are eligible for adaptation. During this step a functional dependence check is carried out to make sure that only those service-components that are compatible for composition are selected. Not only functional components are selected but also MAPE-loop components and, in particular, Planners.
- Secondly, a module called Context Reasoner selects a Planner best matching the current context. This is done by considering how much the "reference context" of each planner matches the actual context.
- The selected Planner then takes over control and makes use of the data acquired in the "M" step and the analyses carried out in the "A" step to propose an adaptation plan.
- Finally, the plan is executed, which results in a parametric and structural adaptation of the whole application.

By considering the metrics defined in Section 2, we observe what follows:

- As it is the case in "conventional" MAPE loop systems, also in ACCADA adaptation is **parametric and structural** and makes use of **predictive individual behaviours**.
- Context-awareness is central in that it determines both adaptation and meta-adaptation.
- Specific and explicit support for a wide range of **Perception**, **Analysis**, and **Planning** behaviours is foreseen.
- **Dynamicity** is medium level: a dynamic trade-off among Perception, Analysis, and Planning is continuously sought. At the same time, an optimal system-environment fit is also continuously sought by selecting a Planner best matching the current contextual conditions.

In ACCADA, formal models may be used to verify and validate certain application configurations. Resilience may then be sought, e.g., by reverting from unresilient configurations to the "known" ones. Another option is to checkpoint application configurations and to roll back to previous configurations when identity losses are detected. Complexity is considerably increased with respect to "plain" MAPE-loop systems, but the service orientation of OSGi (Anonymous, 2009) allows some portion of said complexity to be made transparent to the programmer.

### 3.2.3 Transformer

Transformer is an evolution of ACCADA and, as ACCADA, it is a software framework for the meta-adaptation of service-components MAPE-loop applications (Gui et al., 2013). Also Transformer is based on OSGi 4.0 (Anonymous, 2009). The specific difference of Transformer with respect to its predecessor is given by the introduction of the concept of Planner Fusion. Through Planner Fusion, several Planners best matching the environmental conditions are selected for execution instead of a single one as it is done in ACCADA. The adaptation plans produced by all the selected Planners are then merged together and produce a "global adaptation plan" which is then executed so as to steer parametric and structural adaptation.

Adaptation plans merging is a complex procedure that requires the definition of a formal semantics for a user-defined set of adaptation actions—typically to enable and disable service-components and to set their parameters. This allows conflicts among actions to be formally and semantically defined and detected. Once a conflict is uncovered, a conflict resolution procedure is started. Such procedure takes the set of conflicting actions and either selects a single action out of the set (which is called "winner-takes-all") or synthesises a new action through a sort of "weighted averaging" of the proposed actions. In the latter case the weights are the fitness percentage of the corresponding Planners with respect to the current context. The cited reference provides more detail on Transformer, its design and implementation, and some case studies showing the effectiveness of Transformer as framework for the self-adaptation of component-based applications.

Once again we make use of the metrics defined in Section 2 to draw a number of observations:

- Most of the observations we made for ACCADA hold for Transformer as well; in particular, also in Transformer adaptation is **parametric and structural** and based on **predictive individual behaviours**; context-awareness is key as it steers both adaptation and meta-adaptation. Furthermore, Transformer inherits some aspects of ACCADA's strategy of continuous re-adaptation aimed at reaching an optimal system-environment fit.
- **Perception** and **Awareness** behaviours are as in ACCADA.
- **Planning** and **Dynamicity** is where the major differences are. As already mentioned, Transformer operates a multiple selection of Planners followed by procedures of conflict detection and resolution. Instead of selecting an existing Planner as in ACCADA, Transformer basically synthesises a new composite "Virtual Planner" thus reaching maximal **Dynamicity**. The synthesised new Planner is virtual in that its plan is the result of run-time selections / weighted averages of the actions of the selected Planners. As such, said plan is unknown at design time. All the considerations made for ACCADA with reference to methods to guarantee the identity of the system do not apply anymore in Transformer. In fact it is even possible that Transformer be caught in "adaptation loops", i.e., unbound oscillations between to adaptation states. Because of this we introduced mechanisms to detect and break those adaptation loops. Obviously such mechanisms do not guarantee full coverage from losses of identity, for the absence of adaptation loops is no guarantee of the correctness and the resilience of an adaptation plan. A promising solution to this problem may be found in the approach described in (Lúcio and Syriani, 2013)—a formal method that, given a set of requirements and a model known to satisfy those requirements, guarantees that incremental adjustments to said model, if done according to well-defined rules, still satisfy the requirements.

Finally, we remark how in Transformer algorithmic complexity is considerably increased with respect to "plain" MAPE-loop systems and slightly so with respect to ACCADA.

### 3.2.4 Preliminary Conclusions

We have briefly introduced three MAPE-loop systems of increasing complexity. Here we briefly summarize their characteristics by means of the base of properties introduced in Sect. 2.

Table 3. (Behaviour, Dynamicity) couples and other ancillary properties of the three systems sketched in Section 3.2.

| | Perception → | Awareness → | Planning → | Complexity | Predictability | Assumptions Dependence |
|---|---|---|---|---|---|---|
| MAPE | (Purposeful, Low) | (Predictive, Low) | (Purposeful / teleological, Low) | Low | High | Very strong |
| ACCADA | (Purposeful, Medium) | (Predictive, Medium) | (Teleological, Medium) | Medium | Medium | Medium |
| Transformer | (Purposeful, Medium) | (Predictive, Medium) | (Teleological, High) | High | Low | Low? |

All systems are characterized by configurable Perception. Awareness is in all cases predictive and configurable. The major differences regard Planning: in fact, in MAPE-loop systems the Planner is statically defined at compile-time; in ACCADA, context-awareness is used to select at run-time a Planner out of those available in a registry; in Transformer, a "Virtual Planner" is synthesised by merging the plans of a plurality of Planners matching the current context. Such differences call for an increasing amount of algorithmic complexity. The introduced complexity impacts on Predictability (Jen, 2004). The increased ability to adjust the Planning step to the actual environmental conditions reduces Assumptions Dependence in ACCADA and, possibly, more so in Transformer, especially if formal methods such as those in (Lúcio and Syriani, 2013) would be integrated in the Fusion process.

## 3.3 Collective Perception and Awareness: Social Organisations

In what follows we discuss collective forms of Perception and Awareness and their impact on resilience in social organisations. Our focus is the domain of ambient assisted living (AAL). Once again we briefly review three cases: Smart-houses and Tele-care, the Mutual Assistance Community, and the Fractal Social Organisation.

### 3.3.1 Smart-houses and Tele-care Organisations

As well known, the share of the total population older than 65 is constantly increasing worldwide (Anonymous, 2011; Anonymous 2012). This new environmental condition is bringing to the foreground the many shortcomings of traditional healthcare organisations. Though effective when the context was different and a large amount of resources were available to manage a smaller demand, the traditional approach is now becoming too expensive and thus unacceptable. Merely expanding the current organisations without properly rethinking them is simply not working anymore, while persisting in the "old ways" threatens the ability of modern societies to guarantee care for all. As an answer to this need, novel forms of assistance and care are being sought. Information and communication technology is more and more used to compensate for the ever diminishing human-based assistance. Special, wearable or ambient sensors are used to achieve Perception and get insight into the dynamic conditions of patients and elderly persons. Awareness is employed to identify trends and anticipate potentially dangerous situations. Planning is mostly predefined and typically consists of simple actions such as raising alarms and requiring the intervention of healthcare professionals and traditional healthcare organisations such as, e.g., hospitals. Often the Perception mechanisms are often embedded on the patient and in his/her house, which is then referred to as a so-called "smart house". Awareness and Planning are mostly located elsewhere, in a "tele-care centre". The two sides of the system operate through possibly redundant communication channels, which transfer the Perception data either continuously or at predefined time slots. This leads to the name of "tele-care" for this class of systems.

By considering the metrics defined in Section 2, we observe what follows:

- Tele-care consists of the union of three "organ systems"—one for Perception (the smart-house), one for Awareness and Planning (the tele-care centre), and one for Execution (the traditional healthcare organisation, e.g., a hospital, which performs its classic service when an alarm is triggered by the tele-care centre).
    - The smart-house is a simple **teleological behaviour system**: its function is simply to check continuously whether a given risk threshold is reached. When this is the case, the system "fires", i.e., alerts the tele-care centre. As such, it may be considered as a sort of "macro-sensor" for the tele-care centre. In this sense we shall say that the organisation's **Perception is teleological**.
    - As for **Awareness**, we observe how the tele-care centre is a **predictive behaviour system**: if the data received from the smart-house include a "fire" event, the tele-care issues an alarm to the next level, otherwise it collects the data and analyses them. Analysis here usually includes those of human professional care-givers. If any of the analyses reveal possibly dangerous conditions, the tele-care

centre issues an alarm to the next level, otherwise the data and the extracted knowledge are logged to be used in future analyses.
    - With reference to **Planning**, the healthcare organisation is a **purposeful, non-teleological behaviour system**, in that it is meant to respond with standard, predefined care-delivery procedures to external events (care-requests).
- Dynamicity is low in all resilience components as procedures and protocols are usually predefined statically at design time.

Other, more general considerations are as follows:

- Resilience depends on the truth value of design assumptions and in particular on the ability of the organ systems to rapidly communicate their data and states.
- Resilience and performance depend on the Perception coverage of the smart-house; on the Awareness coverage of the tele-care centre; and on the resilience of the healthcare organisation.
- The algorithmic complexity introduced by the smart-house and the tele-care organisation is limited, statically predefined, and known at design time.
- The operational cost of the system is a function of the number of false positives issued. A false positive in fact results in a costly intervention of the traditional healthcare organisation, involving several costly resources such as ambulances, care teams, professional carers, and so forth.

### 3.3.2 The Mutual Assistance Community

A common aspect shared by most tele-care organisations is that people are divided into classes: primary users, i.e., elderly or impaired people in need of assistance; secondary users (professional providers of care, e.g., doctors and nurses); and tertiary users (society at large). This artificial classification limits the effectiveness of optimally recombining the available assets into an effective and timely response to requests for assistance. Furthermore, this classification into an active part of society, able to contribute with worthy services, and a "passive" part only on the receiving side is already a source of discomfort for people that are thus brought to feel they were once part of a society that now confines them to a lesser state and dignity. The Mutual Assistance Community (MAC) is a social organisation that avoids such classification. In a MAC, the users are just members of a community—they may be, e.g., the citizens of a small village or a group of persons who subscribed to the same gym course. Members are *diverse*, and this translates into a rich variety of services. Diversity implies here different know-how's (as an example, those of a general practitioner, or those of a retired professor of biology); different service policies (for instance, well defined time schedules and associated costs, or "best-effort" voluntary services as occasional informal carers); different value systems, and so on. Another important feature of members is that their "mobile" character: members would move around getting dynamically closer to or farther from other members. Members are not just human beings but also cyber-physical systems, e.g., an accelerometer or a wearable sensor network, or even social organisations—for instance, a hospital.

When a request for assistance is issued, the MAC orchestrates a response by considering the available members; their competence; their availability; and their location with respect to the requester. A key aspect here is the ability to reason in an intelligent way about the nature of the requests and that of the available assets. By unravelling analogies through semantic reasoning we proved that it is possible to reach both a higher level of resource utilisation and a stronger degree of eInclusion (Sun et al., 2007a).

In a MAC, elderly people would not always be passive receivers of care and assistance, but occasionally they could play active roles. As an example, if member *A* feels lonely and wants to have a walk with someone, while member *B* feels like having some physical exercise with someone, then the MAC is able to capture the semantic similarities of those requests and realize that *B* could be the care-giver of *A* and vice-versa—through the so-called "participant" model of our system (Sun et al., 2007a). Societal resources can then be spared, at the same time also preserving human dignity. Simulation (Sun et al., 2006) indicated that systems such as this—able that is to intelligently exploit the dynamically available resources—have the potential to reduce significantly societal costs at the same time increasing efficiency, manageability, and social inclusion.

The MAC is organized into a star topology: a number of peripheral units (members and resources) are connected to a so-called Service Coordination Centre (SCC). The SCC manages a registry similar in purpose to the service registry of service-oriented architectures. The Registry includes information on the members, which is continuously updated, and requests for assistance. Requests are managed by looking for members willing to play specific roles—for instance, if an elderly person falls, an accelerometer member may issue a request for intervention requiring a general practitioner, a nurse, and an ambulance. New and unfulfilled requests (namely, requests for which one or more executive roles are missing) enter the registry and are queued; at each new change of state, semantic methods are invoked to verify whether the current members in their current state may now fulfill roles that previously were missing; if this step fulfills the requirements of one or more unfulfilled requests, the corresponding members are alerted. Corresponding roles are proposed to the members, and if they are accepted the request is marked as fulfilled. More information on the MAC may be found in (Sun et al., 2007b; Sun et al., 2010; De Florio et al., 2013).

By considering the resilience metrics defined in Section 2 we may observe what follows:

- The MAC consists of the union of two organ systems—the "peripheral" system (smart-houses, cyber-physical systems, informal care-givers, professional care-givers, etc.), which manages the Perception and local Planning steps, and the SCC, which centralises the Analysis and the global Planning services.
    - The peripheral system includes a wide diversity of members—hence a wide range of possible behaviours. In practice, **Perception**-oriented members mostly exhibit **teleological behaviours** (as explained in Sect. 3.3.1 for smart-houses) while local **Planning** members may produce up to **resilient social behaviours** depending on the situation at hand and their social and emotional involvement. **Dynamicity** is high due to the collective nature of the system.
    - SCC exhibits two types of behaviours: **predictive individual behaviours** in the Awareness step, in which the "big data" produced by the peripheral systems is semantically correlated. Global Planning is considerably simpler as it only includes **purposeful, non-teleological behaviours**. **Dynamicity** is low as actions are static and predefined.

As we did in previous subsections, we now add a few general considerations:

- The MAC is a centralised organisation. This reduces considerably complexity but affects other properties, e.g. scalability, performance, and predictability. The SCC and the registry are both single-points-of-failure: failures would immediately result in losses of resilience[vi].
- The MAC is a best-effort approach. Mechanisms are required to guarantee—to some extent—the quality of service of the members. In particular human members may exhibit contradictory or unexpected behaviours affecting resilience.
- Social resilience and overall performance depend on individual resilience and performance.
- The algorithmic complexity of the MAC is mostly due to the Awareness step, which requires the execution of computationally intensive semantic match procedures (Sun et al., 2007a; Sun et al., 2007b).

### 3.3.3 The Fractal Social Organisation

The Fractal Social Organisation (FSO) is a distributed organisation that aims at extending the scope and some of the limitations of the MAC. Among this class of organisations, particularly interesting in the context of this paper are so-called holarchies and fractal organisations due to their greater resilience and scalability with respect to "traditional" solutions such as the centralised and the hierarchical. Both holarchies and fractal organisations are characterised by a hierarchical architecture built up through the recursive application of a same building block, respectively called *holon* or *fractal*. The key idea of the FSO is to make use of a system similar to the MAC as the building block of a social organisation. Such building block is called service-oriented community (SoC) (De Florio et al., 2013; De Florio and Blondia, 2010). "Simultaneously a part and a whole, a container and a contained, a controller and a controlled" (Sousa et al., 2000), the FSO models social organisations as the blocks in a Russian nested doll. Each block organises the action of members in proximity of location and intent similarly to the way the MAC manages their resources. Activities are triggered by Perception and Awareness so as to deal with ongoing "situations" (Ye et al, 2012), e.g., a fallen elderly woman in her smart-house. As in MAC, activities call for roles. Roles are either resolved within the current "block"—if said block includes a member able and willing to play the requested role—or forwarded to the next higher level block, if such system is available. In the first case a subset of the current block "commits" to deal with the situation at hand. This subset of members socially committed towards reaching a common objective is called a *community*. In the second case, a request to form a community is forwarded to the next higher level block. We call this an "exception". When a role can be found, the originating community becomes a member of a higher level community, otherwise the request fails. Situation treatment protocols are made available through meta-data and the community lifecycle is realized through a modified service-oriented model—which brought to the name "service-oriented community". More information on the FSO and its canon may be found in (De Florio et al., 2013).

As can be seen from the above concise description, the FSO introduces a second "dimension" with respect to the flat, mono-dimensional organisation of the MAC. Multiple, concurrent MAPE-loop systems are present both horizontally (that is, at block level) and vertically (at organisational level). The main benefits of this approach are the greater manageability and scalability with respect to traditional solutions and the flat MAC (due to the modularity and the self-similarity of the FSO hierarchy) as well as a "smarter" use of the available resources compared to the static resource allocation typical of traditional organisations. A typical case of this is a community $C$ within SoC "hospital $A$" that is lacking a certain resource, e.g. a CAT scanner. The missing role triggers the role exception to a higher level, e.g., a "meta hospital $M$". $M$ is in fact a registry of hospitals and their resources, including a "hospital $B$" that has and currently is not using a CAT scanner. A new community is then formed by $C$ and $B$ to deal with the situation that triggered the exception. Once that situation is dealt with, the roles are released and the community disappears.

In what follows once again we apply the resilience metrics defined in Section 2 in order to derive some facts about the FSO.

- The FSO is a hierarchical, distributed social organisation of MAPE-loop systems. MAPE-loop components are present at all levels, which means in particular that Perception and Awareness are collective and distributed—thus reaching the highest degree of **Dynamicity**. Each system is at the same time "peripheral" and "central" (in the sense given to those adjectives in Sect. 3.2.2). The behaviour of Perception and Awareness components is not constrained by fixed and predefined roles. On the contrary, such behaviours may vary with the context and with the role each component commits to play. Strategies are collective and based on collaboration and mutual assistance, which means that **Perception** and **Awareness** are characterised by **resilient social behaviours**. Due to intrinsic (algorithmic) limitations and its fixed role, Planning is only **teleological** and with low **Dynamicity**.

Other more general considerations are as follows:

- Through its hierarchical nature, the FSO avoids some of the negative aspects of the MAC—e.g., the presence of single-points-of-failure. Self-similarity and other morphological properties (De Florio et al., 2013) suggest that the FSO enhances considerably scalability and performance—albeit no actual verification of those properties is yet available. Through its distributed nature, the FSO reduces considerably management complexity. Finally, through its modularity, the FSO greatly enhances its *evolvability* (i.e., its ability to rapidly respond to the changes introduced by turbulent environments (Clune et al., 2013).)
- The FSO is still a best-effort approach, though it wraps straightforwardly existing organisations both traditional and new—for instance, hospitals, tele-care organisations, and MACs. We conjecture that this may result in greater predictability, reliability, and quality of service.
- Social resilience and overall performance are less dependent on individual resilience and performance, as local failures result in role exceptions that may be dynamically tolerated by involving entities visible at higher levels in the FSO hierarchy.
- The algorithmic complexity of the FSO is distributed across a plurality of systems and levels, which implies a high degree of parallelism in all the MAPE-loop components.

### 3.3.4 Preliminary Conclusions

We have briefly introduced three organisations of increasing complexity. Here we briefly summarize their characteristics by means of the base of properties introduced in Sect. 2.

Table 4. (Behaviour, Dynamicity) couples and other ancillary properties of the three systems described in Section 3.4

|  | **Perception →** | **Awareness →** | **Planning →** | Complexity | Predictability | Assumptions Dependence |
|---|---|---|---|---|---|---|
| Smart-house + tele-care | (Teleological, Low) | (Predictive, Low) | (Purposeful, Low) | Low | High | Very strong |
| MAC | (Teleological, High) | (Predictive, Low) | Local aspects: (Resilient social, High) Central aspects: (Purposeful, Low) | Medium | Low | Medium |
| FSO | (Resilient social, High) | (Resilient social, High) | (Teleological, Low) | High | Medium | Low |

As summarised in Table 4, the three surveyed organisations are characterised by remarkable behavioural differences.

- **Perception** is very simple and **teleological** in the smart-house. **Dynamicity** is low due to the predefined set of sensing devices that are deployed. **Perception** is also **teleological** in the MAC, but **Dynamicity** is high due to the multiple perception layers of the MAC members and to the adoption of service-orientation, which allows for full dynamic reconfigurability of the Perception sources. In the FSO, **Perception** goes one step further. In fact Perception data manifesting in one layer, once analysed and processed, may take two different "routes": they may be either resolved in that layer, or forwarded to a higher layer via the exception mechanism, thus becoming new Perception data for that layer. Such collective strategy allows Perception data to be made transparent or translucent depending on the context. As a result of this, the behaviours of the Perception service may be recognized as **resilient social**.
- Similar considerations apply to **Awareness**: in tele-care and the MAC, Awareness takes simple forms of **predictive** behaviour. In FSO, Awareness may be resolved in one layer or go up through the hierarchy involving more and more complex entities (thus resulting in **resilient social behavior**). **Dynamicity** is high for the same reasons we discussed for Perception.
- **Planning** behaviour is very simple in all the three cases. Tele-care organisations simply initiate an alarm procedure—a simple **purposeful** behaviour as in Boulding's "Thermostats". Though slightly more complex as it involves enrolling multiple members into a "social response", Planning in MAC's is also purely **purposeful**. In FSO, Planning needs to consider multiple cases—as many as the layers that need to be traversed in order to activate a social response to a situation detected in one layer. The decision to "sink" the response or forward it to the next layer makes of the Planning step in FSO a case of **teleological behaviour**.

# 4 Conclusions

In this paper we suggested that resilience in systems and organisations may be better understood as a multi-attribute property, "defined and measured by a set of different indicators" (Costa & Rus, 2003). We proposed three such indicators—Perception, Awareness, and Planning—and suggested a bidimensional classification. In this classification, a first coordinate is behavioural and refers to the characteristics of the action of the system components. The second coordinate characterises the stability of those system components and expresses whether they are statically predefined and immutable; or dynamically reconfigurable or selectable; or they are the result of a collective action of a fusion of multiple and different features. In order to demonstrate the expressiveness of our classification we applied it to three families of systems and organisations, each of which includes three members of increasing complexity. Results are depicted in Table 5 and Table 6.

Table 5. (abscissa, ordinate) identifiers respectively corresponding to the behaviours of Perception/Awareness/Planning sub-systems and to their Dynamicity.

| Behaviour | Abscissa | Dynamicity | Ordinate |
|---|---|---|---|
| Passive / none | 1 | Static | I |
| Active, non-purposeful | 2 | Dynamic or selectable | II |
| Purposeful, non-teleological | 3 | Mergeable or collective | III |
| Teleological, non-predictive | 4 | | |
| First-order predictive | 5 | | |
| Second-order predictive | 6 | | |
| n-order predictive, n>2 | 7 | | |
| Collective resilient behaviours | 8 | | |
| Complex collective resilient behaviours | 9 | | |

Table 6. Summary of the resilience characteristics for the systems surveyed in Section 3. Values refer to the identifiers introduced in Table 5.

| Rows 1–9: Systems, Cols 2–4: Resilience sub-systems | Perception → | Awareness → | Planning (L=local, C=central) → |
|---|---|---|---|
| RDS | (1,I) | (3,I) | (3,I) |
| A-RDS | (1,I) | (5,I) | (3,I) |
| A-NVP | (1,I) | (6,I) | (3,II) |
| MAPE | (3,I) | (5–7,I) | (3–4,I) |
| ACCADA | (3,II) | (5–7,II) | (4,II) |
| Transformer | (3,II) | (5–7,II) | (4,III) |
| SH+telecare | (4,I) | (5–7,I) | (3,I) |
| MAC | (4,III) | (5–7,I) | L: (3–9,III),C: (3,I) |
| FSO | (8–9,III) | (8–9,III) | (4,I) |

It is our final conjecture here that tables such as Table 6 may help putting on the foreground the most insightful features of a system with respect to its ability to attain resilience. This paves the way to future approaches were Perception, Awareness, and Planning may be expressed as a "resilience contract" to be matched with an "environment policy"—in the sense discussed in (Eugster et al., 2009). Matching contract with policy in this case would mean to set up a handshake mechanism between a system declaring its resilience figures and an environment stating the expected minimal resilience requirements. Breaking down resilience into its main constituents allows a fine-tuning of the necessary dynamic trade-offs while reducing at the same time the complexity and energy budgets necessary to back up the selected system architectures. This principle could be used to set up *meta-resilient systems*, namely systems and organisations able to (i) introspect about their own resilience attributes; (ii) analyse the probability of shortcoming or excess in Perception, Awareness, and Planning; and (iii) self-optimise themselves accordingly. This would allow a system to adopt a resilience approach minimising the chances that the resulting architecture be either insufficiently structured to deal with the mission at hand or not enough "gifted" to guarantee persistence of its features—a concept similar to the autonomic minimisation of overshooting and undershooting introduced in Sect. 3.1 for the optimal management of redundancy.

## Acknowledgements

We would to express our gratitude to the reviewers and acknowledge how their suggestions and remarks considerably contributed to the quality of this work.

## References

(Adner & Kapoor, 2010)
　　Adner, R. and Kapoor, R. Value creation in innovation ecosystems: How the structure of technological interdependence affects firm performance in new technology generations. *Strategic Management Journal*, 31, pp. 306–333.
(Anonymous, 2009)
　　Anonymous. *OSGi Service Platform, Core Specification, Release 4, Version 4.2*. OSGi Alliance, 2009, pp. 332, ISBN 978-90-79350-04-9.
(Anonymous, 2011)
　　Anonymous. The 2012 ageing report: Underlying assumptions and projection methodologies. *European Economy* 4, 309 (2011). DOI 10.2765/15373
(Anonymous, 2012)
　　Anonymous. Population structure and ageing. Tech. rep., Eurostat (2012).
　　URL http://epp.eurostat.ec.europa.eu/statistics_explained/
(Astley & Fombrun, 1983)

Astley, W. G., and Fombrun, C. J. Collective Strategy: Social Ecology of Organisational Environments. *The Academy of Management Review*, 8(4), Oct. 1983, pp. 576–587.

(Aristotle, 1986)

Aristotle, 1986. *De anima (on the soul)*. Translation and notes by H. Lawson-Tancred. Penguin classics. Penguin Books.

(Avižienis, 1985)

Avižienis, Algirdas. The *N*-version Approach to Fault-Tolerant Software. IEEE Trans. Software Eng. **11**, Dec. 1985, pp. 1491–1501.

(Boulding, 1956)

Boulding, Kenneth. 1956. General systems theory-the skeleton of science. *Management science*, **2**(3).

(Brandenburger & Nalebuff, 1999)

Brandenburger, A. and Nalebuff, B. *Co-opetition—A Revolutionary Mindset that Combines Competition and Cooperation*. Doubleday, 1998, ISBN: 9780385479509

(Buys et al., 2011)

Buys, Jonas, De Florio, Vincenzo, & Blondia, Chris. 2011. Towards context-aware adaptive fault tolerance in soa applications. *Pages 63-74 of: Proceedings of the 5th ACM international conference on distributed event-based systems (DEBS)*. Association for Computing Machinery, Inc. (ACM).

(Buys et al., 2012)

Buys, Jonas, De Florio, Vincenzo, & Blondia, Chris. 2012. Towards Parsimonious Resource Allocation in Context-Aware N-Version Programming. *Proceedings of the 7th IET System Safety Conference*. The Institute of Engineering and Technology.

(Charette, 2011)

Charette, Robert. 2011 (January). *Electronic devices, airplanes and interference: Significant danger or not?* IEEE Spectrum blog "Risk Factor", available at URL http://spectrum.ieee.org/riskfactor/aerospace/aviation/electronic-devices-airplanes-and-interference-significant-danger-or-not.

(Clune et al., 2013)

Clune, J. Mouret, J.-B., and Lipson. H. The evolutionary origins of modularity. *Proc R Soc B* 2013 280: 20122863.

(Costa & Rus, 2003)

Costa, Patricia, & Rus, Ioana. 2003. Characterizing software dependability from multiple stakeholders perspective. *Journal of software technology*, **6**(2).

(De Florio et al., 2000)

De Florio, Vincenzo, Deconinck, Geert, & Lauwereins, Rudy. 2000. An algorithm for tolerating crash failures in distributed systems. *Pages 9-17 of: Proc. of the 7th Annual IEEE International Conference and Workshop on the Engineering of Computer-based Systems (ECbS)*. Edinburgh, Scotland: IEEE Comp. Soc. Press.

(De Florio, 2010)

De Florio, Vincenzo. 2010. Software assumptions failure tolerance: Role, strategies, and visions. *Pages 249-272 of:* Casimiro, Antonio, de Lemos, Rogério, & Gacek, Cristina (eds), *Architecting dependable systems vii*. Lecture Notes in Computer Science, vol. 6420. Springer Berlin / Heidelberg. 10.1007/978-3-642-17245-8_11.

(De Florio, 2011a)

De Florio, Vincenzo. 2011a (May). *Imec academy / sset seminar "Cost-effective software reliability through autonomic tuning of system resources."* Available online through http://mediasite.imec.be/mediasite/SilverlightPlayer/Default.aspx?peid=a66bb1768e184e86b5965b13ad24b7dd

(De Florio, 2011b)

De Florio, Vincenzo. 2011b. Robust-and-evolvable resilient software systems: Open problems and lessons learned. *Pages 10-17 of: Proceedings of the 8th workshop on Assurances for Self-adaptive Systems*. ASaS '11. Szeged, Hungary: ACM.

(De Florio, 2011c)

De Florio, Vincenzo. Software Assumptions Failure Tolerance: Role, Strategies, and Visions. Chapter in *Architecting Dependable Systems*, 7, (Antonio Casimiro, Rogério de Lemos, and Cristina Gacek, Eds.), Lecture Notes in Computer Science., Addison-Wesley, 2011.

(De Florio, 2012a)

De Florio, Vincenzo. 2012a (August). On the role of perception and apperception in ubiquitous and pervasive environments. *Proceedings of the 3rd workshop on Service Discovery and Composition in Ubiquitous and Pervasive Environments (SUPE'12)*.

(De Florio, 2012b)

De Florio, Vincenzo. Orchestrating Democracy. Open Democracy, October 10, 2012. Available online at http://www.opendemocracy.net/vincenzo-de-florio/orchestrating-democracy

(De Florio & Blondia, 2007)
    De Florio, Vincenzo, & Blondia, Chris. 2007 (August). Reflective and refractive variables: A model for effective and maintainable adaptive-and-dependable software. *In: Proc. of the 33rd Euromicro Conference on Software Engineering and Advanced Applications (SEAA 2007).*

(De Florio & Blondia, 2008a)
    De Florio, Vincenzo, & Blondia, Chris. 2008. On the requirements of new software development. *International Journal of Business Intelligence and Data Mining*, **3**(3).

(De Florio & Blondia, 2008b)
    De Florio, Vincenzo, & Blondia, Chris. 2008. A Survey of Linguistic Structures for Application-level Fault-Tolerance. *ACM Computing Surveys*, 40(2), 2008.

(De Florio & Blondia, 2010)
    De Florio, Vincenzo, & Blondia, Chris. 2010. Service-oriented communities: Visions and contributions towards social organisations. *Pages 319-328 of:* Meersman, Robert, Dillon, Tharam, & Herrero, Pilar (eds), *On the move to meaningful internet systems: Otm 2010 workshops*. Lecture Notes in Computer Science, vol. 6428. Springer Berlin / Heidelberg. 10.1007/978-3-642-16961-8_51.

(De Florio et al., 2000)
    De Florio, Vincenzo, Deconinck, Geert, & Lauwereins, Rudy. 2000. An algorithm for tolerating crash failures in distributed systems. *Pages 9-17 of: Proc. of the 7th Annual IEEE International Conference and Workshop on the Engineering of Computer-based Systems (ECbS)*. Edinburgh, Scotland: IEEE Comp. Soc. Press.

(De Florio et al., 2013)
    De Florio, V., Coronato, A., Bakhouya, M., and Di marzo, G. 2013. Fractal Social Organisations—Models and Concepts for Socio-technical Systems. Submitted to *Computing*, Springer, 2013.

(Eugster et al., 2009)
    Eugster, Patrick Th., Garbinato, Benoit, & Holzer, Adrian. 2009. Middleware support for context aware applications. Chap. 14, pages 305-322 of: Garbinato, B., Miranda, H., & Rodrigues, Lus (eds), *Middleware for network eccentric and mobile applications*. Springer.

(Gui et al., 2011a)
    Gui, Ning, De Florio, Vincenzo, Sun, Hong, & Blondia, Chris. Toward Architecture-based Context-Aware Deployment and Adaptation. *Journal of Systems and Software*, **84**, Elsevier, 2011.

(Gui et al., 2011b)
    Gui, Ning, De Florio, Vincenzo, & Blondia, Chris. 2011. Run-time compositional software platform for autonomous NXT robots. *International Journal of Adaptive, Resilient and Autonomic Systems,* special ADAMUS issue, **2**(2).

(Gui et al., 2013)
    Gui, Ning, De Florio, Vincenzo, & Holvoet, Tom. Transformer: an adaptation framework with contextual adaptation behavior composition support. Accepted for publication in *Software: Practice & Experience*, Wiley.

(Guseva, 2012)
    Guseva, K. *Formation and Cooperative Behavior of Protein Complexes on the Cell Membrane*. Springer, Springer Theses, ISBN 978-3-642-23987-8. Originally published as a doctoral thesis of the Institute of Complex Systems and Mathematical Biology of the University of Aberdeen, UK

(Harris, 2005)
    Harris, Cyril M. (ed.). *Dictionary of Architecture and Construction*. McGraw-Hill.

(Jen, 2004)
    Jen, Erica. 2004. Stable or robust? What's the difference? *Pages 7-20 of:* Jen, Erica (ed), *Robust design: a repertoire of biological, ecological, and engineering case studies*. SFI Studies in the Sciences of Complexity. Oxford University Press.

(Johnson, 1989)
    Johnson, Brian W. *Design and Analysis of Fault-Tolerant Digital Systems*. Addison-Wesley, New York, 1989.

(Kephart & Chess, 2003)
    Kephart, Jeffrey O., & Chess, David M. 2003. The vision of autonomic computing. *Computer*, **36**(January), 41-50.

(Laprie, 1985)
    Laprie, Jean-Claude. 1985. Dependable computing and fault tolerance: Concepts and terminology. *Pages 2-11 of: Proc. of the 15th Int. Symposium on Fault-tolerant Computing (FTCS-15)*. Ann Arbor, Mich.: IEEE Comp. Soc. Press.

(Laprie, 1995)
    Laprie, Jean-Claude. 1995. Dependability-its attributes, impairments and means. *Pages 3-18 of:* Randell, B., Laprie, J.-C., Kopetz, H., & Littlewood, B. (eds), *Predictably dependable computing systems*. ESPRIT Basic Research Series. Berlin: Springer Verlag.

(Lúcio and Syriani, 2013)
> Lúcio, Levi and Syriani, Eugene. Assuring quality in iterative modeling through property preservation. Unpublished document. Available at http://property-preserving-evolution.googlecode.com/svn-history/r22/trunk/trunk/Property_oriented_model_evolution/assisted_model_evolution.pdf

(Meyer, 2009)
> Meyer, John F. 2009 (Sept.). Defining and evaluating resilience: A performability perspective. *In: Proceedings of the International Workshop on Performability Modeling of Computer and Communication Systems (PMCCS).*

(Nilsson, 2010)
> Nilsson, Thomy. How neural branching solved an information bottleneck opening the way to smart life. *In: Proceedings of the 10th International Conference on Cognitive and Neural Systems.*

(Nilsson, 2008)
> Nilsson, Thomy. 2008. Solving the sensory information bottleneck to central processing in complex systems. *Pages 159-186 of:* Yang, Ang, & Shan, Yin (eds), *Intelligent complex adaptive systems*. Hershey, PA: IGI Global.

(Rosenblueth et al., 1943)
> Rosenblueth, Arturo, Wiener, Norbert, & Bigelow, Julian. 1943. Behavior, purpose and teleology. *Philosophy of science*, **10**(1), 18-24.

(Sachs, 1995)
> Sachs, Joe. 1995. *Aristotle's Physics: A guided study*. Masterworks of Discovery. Rutgers University Press.

(Salehie & Tahvildari, 2009)
> Salehie, Mazeiar, & Tahvildari, Ladan. 2009. Self-adaptive software: Landscape and research challenges. *ACM Trans. Auton. Adapt. Syst.*, **4**(May), 1-42.

(Schultz et al., 2009)
> Schultz, D., Wolynes, P. G., Ben Jacob, E., and Onuchic, J. N. Deciding Fate in Adverse Times: Sporulation and Competence in *Bacillus subtilis*. *Proc. Nat.l Acad. Sci.* 106, pp. 21027–21034, 2009.

(Sousa et al., 2000)
> Sousa, P., Silva, N., Heikkila, T., Kallingbaum, M., Valcknears, P.: Aspects of co-operation in distributed manufacturing systems. *Studies in Informatics and Control Journal* **9**(2), 89–110 (2000)

(Stephenson et al., 2010)
> Stephenson, A., Vargo, V., and Seville, E. (2010). Measuring and Comparing Organisational Resilience in Auckland. *The Australian Journal of Emergency Management*, 25(2), 27-32.

(Sun et al., 2006)
> Sun, Hong, De Florio, Vincenzo, & Blondia, Chris. 2006. A design tool to reason about ambient assisted living systems. *In: Proceedings of the International Conference on Intelligent Systems Design and Applications (ISDA'06)*. Jinan, Shandong, P. R. China: IEEE Computer Society.

(Sun et al., 2007a)
> Sun, Hong, De Florio, Vincenzo, Gui, Ning, & Blondia, Chris. 2007. Participant: A new concept for optimally assisting the elder people. *In: Proceedings of the 20th IEEE International Symposium on Computer-based Medical Systems (CBMS-2007)*. Maribor, Slovenia: IEEE Computer Society.

(Sun et al., 2007b)
> Sun, Hong, De Florio, Vincenzo, Gui, Ning, & Blondia, Chris. Service Matching in Online Community for Mutual Assisted Living. In: *Proc. of the 3rd International Conference on Signal-Image Technology & Internet based Systems* (SITIS' 2007), Shanghai, China, Dec. 2007.

(Sun et al., 2010)
> Sun, Hong, De Florio, Vincenzo, Gui, Ning, & Blondia, Chris. The Missing Ones: Key Ingredients Towards Effective Ambient Assisted Living Systems. *Journal of Ambient Intelligence and Smart Environments*, **2** (2), April 2010, pp. 109–120.

(Ye et al. 2012)
> Ye, J., Dobson, S., McKeever, S. Situation identification techniques in pervasive computing: A review. *Pervasive and Mobile Computing* **8**(1), 36–66 (2012)

(Webb & Levin, 2004)
> Webb, Colleen T., & Levin, Simon A. 2004. Cross-system perspectives on the ecology and evolution of resilience. *Pages 151-172 of:* Jen, Erica (ed), *Robust design: a repertoire of biological, ecological, and engineering case studies*. SFI Studies in the Sciences of Complexity. Oxford University Press.

(Zhuge, 2010)
> Zhuge, H. Cyber Physical Society—A Cross-Disciplinary Science. On-line document. URL: http://www.knowledgegrid.net/~h.zhuge/CPS.htm

[i] Vincenzo De Florio is a post-doctoral researcher with the PATS research group of the University of Antwerp. There he is responsible for the PATS' task force on adaptive-and-dependable software systems, namely software, devices, and services that are built so as to sustain an agreed-upon quality-of-service and quality-of-experience despite the occurrence of potentially significant and sudden changes or failures in their infrastructures and environments. Vincenzo published 3 books (one as author, two as editor), 23 journal papers, 3 book chapters, and 77 conference papers. His H-index is 13 (Google Scholar). He is or was member of several program committees for international conferences and workshops such as PDP (International Conference on Parallel, Distributed, and Network-Based Processing), ARES (the International Conference on Availability, Reliability and Security, SSDU (the International Workshop on Service, Security and its Data management technologies in Ubi-com), and DASC (the IEEE International Symposium on Dependable, Autonomic and Secure Computing). He was co-founder and co-organiser of the international workshop ADAMUS (on Adaptive and DependAble Mobile Ubiquitous Systems), for which he also served as co-chair in its third and fifth editions. He is co-inventor of a European/US patent. Vincenzo is co-proposer and principal investigator of project ARFLEX (Adaptive Robots for FLEXible manufacturing systems, IST-NMP2-016680) and of iMinds project "Little Sister" (started in January 2013). Vincenzo is the Editor-in-chief of IJARAS, the International Journal of Adaptive, Resilient and Autonomic Systems.

[ii] From Greek word "ataraxy", namely the quality of being able to proceed without minding any external event or condition; from *a-*, not, and *tarassein*, to disturb.

[iii] By this we mean here to the strategy through which the conventional memory cells associated to a redundant cell are allocated in main memory. Sequential allocation or fixed-stride allocation are common strategies.

[iv] The term "frozen ducks" was introduced in (De Florio, 2011a) to refer to the fact that a system's specifications, including, e.g., the choice of $N$ in RDS, apply to some possibly distant time and result in that system being fixated ("frozen") on the validity of possibly stale assumptions. The "frozen ducks" syndrome is likely to affect close-world systems when re-deployed or subjected to mutating environmental conditions. A typical case of frozen ducks is efficaciously reported by U.S. engineer Bill Strauss: "A plane is designed to the right specs, but nobody goes back and checks if it is still robust" (Charette, 2011).

[v] In (De Florio, 2012b) we conjectured that a similar result may apply also to collective systems of people organized as democracies.

[vi] Obviously resilience methods, e.g. software elasticity, may be applied to reduce the extent of this critical dependency. A method for this is described for instance in (De Florio et al., 2000).